\documentclass{article}

\usepackage{arxiv}

\usepackage[utf8]{inputenc} 
\usepackage[T1]{fontenc}    
\usepackage{hyperref}       
\usepackage{url}            
\usepackage{booktabs}       
\usepackage{amsfonts}       
\usepackage{nicefrac}       
\usepackage{silence}
\usepackage{lipsum}		
\usepackage{graphicx}
\usepackage{amsmath}
\usepackage{enumitem}

\title{The cooperation-defection evolution on social networks}

\author{ \href{https://orcid.org/0000-0001-8978-4339}{\includegraphics[scale=0.06]{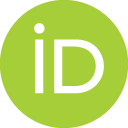}\hspace{1mm} Bijan Sarkar}\thanks{E-mail: bijan0317@gmail.com} \\
	Department of Mathematics \\
	Neotia Institute of Technology, Management and Science \\
	Diamond Harbour Road, 24 Parganas (South) \\
	West Bengal-743368, India \\
	\texttt{bijan0317@yahoo.com} \\
}

\hypersetup{
pdftitle={The cooperation-defection evolution on social networks},
pdfsubject={q-bio.PE, math.DS},
pdfauthor={Bijan Sarkar},
pdfkeywords={evolutionary game theory, analytical procedure, algorithm method, cooperation enhancement, defection hindrance},
}

\begin{document}
\maketitle

\begin{abstract}
Without contributing, defectors take more benefit from social resources than cooperators which is the reflection of a specific character of individuals. However, natural physical mechanisms of our society promote cooperation. Thus, in the long run, the  evolution about genetic variation is something more than the social evolution about fitness. The loci of evolutionary paths of the cooperation and the defection are correlated, but not a full complement of each other. Yet, the only single specific mechanism which is operated by some rules explains the enhancement of cooperation where the independent analysis of defect evolutionary mechanism is ignored. Moreover, the execution of a particular evolutionary rule through algorithm method over the long time encounters highly sensitive influence of the model parameters. Theoretically, biodiversity of two types relatively persists rarely. Here I describe the evolutionary outcome in the demographic fluctuation. Using both analytical procedure and algorithm method the article concludes that the intratype fitness of individual species is the key factor for not only surviving, but thriving. In consideration of the random drift, the experimental outcomes show that dominant enhancement of cooperation over defection is qualitatively independent of environmental scenario. Collectively, the set of the rules  becomes an evolutionary principle to cooperation enhancement. 
\end{abstract}


\section{Introduction}

The demographic fluctuation is not the consequence of a particular account of evolutionary structure\cite{houchmandzadeh2015fluctuation, huang2015stochastic, mcavoy2018public}. It is partly an effect of the random fluctuation in the result of the combination of various genetic, ecological and sociological  factors such as selection, reproduction, mutation, migration, etc. The factors collectively play a role in the struggle for existence \cite{mallet2012struggle}. Thus, a
 random drift on the evolutionary structure would be needed to explain the stochastic effect of demographic fluctuation. As the clarification of this effect can be based on some logical structure, it obviously has deterministic interpretation. Here, we explore the random fluctuation features by adopting the concept of individuals' attachment and leaving tendency to a social network. It has generally found that any kind of regularity break in the social structure pattern in the sense of either geometry or individual reputation and punishment help to cooperation enhancement. However, it is not true in all circumstances at all \cite{hauert2004spatial, kleineberg2017metric}.

Over the evolutionary game theory framework the birth-death evolution can robustly define the  evolutionary dynamics of pure Darwinian selection~\cite{champagnat2006unifying,simon2008stochastic,doebeli2017point}. To account for this emergence of cooperation it is primarily emphasised on the configurational arrangement of individuals in their residential areas where the total population size is always fixed over time. Assuming so, the derived rules of the analytically tractable model, based on the formalisms like Hamilton's rule \cite{ohtsuki2006simple,nowak2006five,rand2013human} as well as the static game property like Evolutionarily Stable Strategy \cite{nowak2004emergence,santos2008social}, are able to figure out the ubiquitous character of cooperation. However, every rule has its own limitation, and they fail over long ranges of biological parameters~\cite{van1998unit}. Thereby, instead of emergence principle the emergence theory has been compelled to rely on some specific rules of limiting evolutionary cases, up until now. It has found, generally, the structured complex networks defined by either the hub unit or the cluster unit support the cooperation enhancement. However, in unbiased initial distribution of individuals it is still not well understood that, what type of driving force is involved so that cooperators can take advantage of complex network whereas defectors fail to take same advantage.

In the population genetics, the Moran process and the Wright-Fisher process describe how phenotypic variation occurs over time by evolution. The working procedure of both processes is dependent on the individual fitness. To ensure the population size remains constant in each time step of the Moran process a random individual is chosen proportional to its fitness for reproduction and another random individual of less fitter is chosen for death. The Wright-Fisher process work as synchronous concept where the possibility of the entire population to be replaced by the identical offspring from previous generation is given by the binomial sampling probability. Here, we utilise the concepts of both processes simultaneously. The birth and death selection is governed by the Moran process and the binomial sampling probability determines the fertility effect. The combined processes accounts for the fertility effect and the viability effect at the same time \cite{scott1988conditions}. However, in the algorithm version of our hybrid system we are able to determine the segregation of genes by the simple iterative method instead of the binomial distribution.   

The presence of graphs of connections among finite individuals is indicative of the qualities for real populations in which cooperation may gradually evolve starting from discrete or fuzzy initial conditions. The evolution of a specific type under the topics of the fixation probability and time taking by a mutant to take over a population on a graph has been investigated in several studies (see \cite{broom2013game} and references therein). Besides the analytical results, simulation as well as numerical approximations are also shared in the same branch of knowledge, but the obtained results are often hard to interpret or far from being a real scenario. The stochastic and the deterministic both equations of evolutionary game on networks of the finite but large population of individuals those are able to switch their types have been introduced to measure the influence of the network topology \cite{traulsen2005coevolutionary, iacobelli2016lumping, sarkar2018moran}. However, most models of evolutionary dynamics are well fitted only for the time equation of the probability distribution of types over a well-mixed infinite population. The contribution of this paper is the introduction of a new model of two versions of a hybrid system for studying evolutionary dynamics.

We already know that the stochastic effects become less important for large population size \cite{traulsen2005coevolutionary}. In the continuation, it is also true that the existence of non Darwinian dynamics  can only expect regarding the random scenarios. Thus, how the local evolution of each of the hub populations affecting on global evolutionary dynamics is one of the crucial branches of evolutionary research work, where the term hub best suits for small, simple neighbourhood area network environment of a focal node.

We consider a social network by a simple graph structure of $\mathcal{N}$ vertices. Every vertex is occupied by a cooperator ($C$), occupied by a defector ($D$) or  vacant -- the place which could be occupied by any one of the two types (or species) of individuals. Two individuals can interact only if they are connected by an edge of the graph. We confine ourselves in on the random regular graphs, however, implementing the model concept for the heterogeneous complex networks is straightforward. The evolution is interpreted here by the genetic reproduction, and the model formalism is well fitted for the both asexual and sexual contexts. Besides the birth-death update event, the synchronous formulation  incorporates the random drift effect while the internal migration is overlooked.

In the social pool, the evolutionary process advances through the function of the fitness values of the individuals, where the fitness measures by the combination of the fertility effect and the viability effect on gene $C$ and gene $D$ \cite{bodmer1965differential,kidwell1977regions}. Effective fitness generally is studied at the reproduction level of an individual \cite{sarkar2016random}. The fitness means the effective fitness here. We consider the standard payoff matrix of a two-player version of the interaction for the representing the fitness values, $\pi_{**}$, of the row individual. According to the base matrix of such interaction, cooperators pay a cost to provide a benefit while defectors neither pay cost nor contribute benefit.

Based on the experimental outcome obtained by using both analytical procedure and algorithm method where the procedure and the method are derived through a trick of the putting mathematical tractability before empirical evidence in this article we infer that the intratype fitness of individual species is the key factor for cooperation enhancement and defection hindrance. A set of five self-explanatory observables clarifies the whole experimental outcome. In following the slightly non-conventional way our aim is to demonstrate that in the distinct mechanisms of cooperation and defection, natural cooperation is a principle of evolution. We strongly believe that a scientific explanation can be accomplished through a simple easier logical procedure, and a time consuming trick can also be replaced by an elegant trick.

\section{The model}
\label{sec:headings}
General knowledge of the underlying interaction nature of the ground  environment is essential to build up an evolutionary model. In the consideration of a two-player version of the interaction, we know that if the payoffs of an $i^{th}$ individual getting from an interaction with a $j^{th}$ individual are denoted by the $\pi_{ij}$, then in general the base payoff matrix for the cooperation-defection interaction is given by: $\pi_{CC}=b-c$, $\pi_{CD}=-c$, $\pi_{DC}=b$, $\pi_{DD}=0$, where cooperators pay cost $c>0$ to provide a benefit $b>c$. Defectors neither  pay cost nor contribute benefit. Besides this payoff matrix, on the well-known three game classes,  namely Dominance game, Coexistance game and Coordination game, other cooperation-defection payoff matrices can also be defined to encompass general social dilemmas between cooperators and defectors where the general behavioural character of cooperator and defector is inherently incorporated with each of the games in the sense that in the intratype competition, the cooperator payoff is higher than the defector payoff $(\pi_{CC}>\pi_{DD})$; on the contrary, in the intertype competition a defector gains higher payoff than a cooperator $(\pi_{DC}>\pi_{CD})$, and to get a higher payoff each type of individuals prefers to interact with a cooperator than a defector $(\pi_{CC}>\pi_{CD} \hspace{.8mm}  \mbox{and} \hspace{.8mm} \pi_{DC}>\pi_{DD})$. In this concept, each of the payoff values for various well-known games is bounded into more restricted ways regarding the three game classes \cite{zukewich2013consolidating}. However, it will be important to note that the occurrence of an expected recognised character of a specific game which is operated by the mathematical inequalities of the payoffs of the social dilemmas is highly influenced by the number of presence individuals of the evolutionary system.

Geometrically, we are interested to study the evolutionary dynamics at a particular region where the particular region is surrounded by similar types of neighbour regions. In each of the regions, the presence of a graph of connections among individuals exhibits the type relation among themselves. As it is to be assumed that the effect of random drifts are different for different regions, the influence of the neighbour regions on evolutionary dynamics at the focal region is of interest. Each region comprises the different sizes of hubs of the graph.

\begin{figure}[t!]
\begin{center}
  \includegraphics[width=0.7\linewidth]{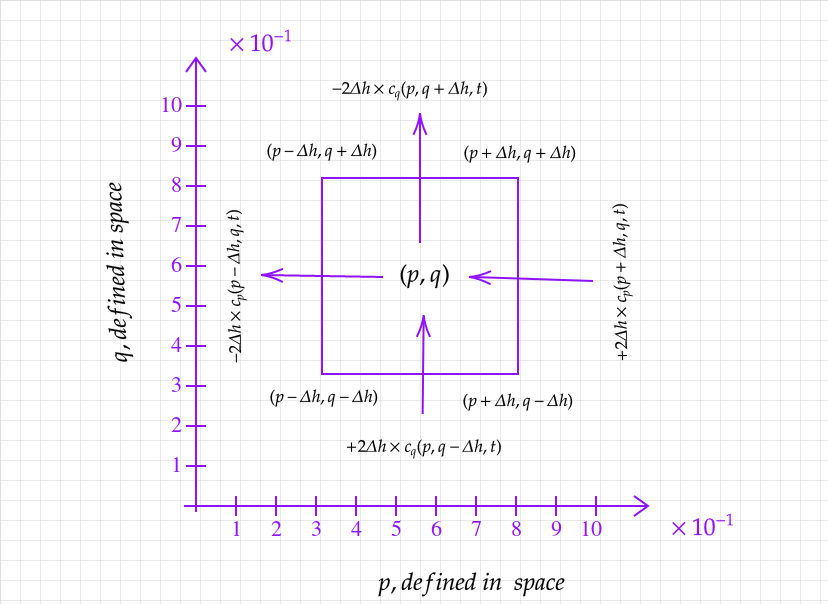}
  \caption{An accumulation of individuals through the random drift in a small area, $(2\Delta h)^{2}$, on the probability space $\Omega$. The $\Omega$ is embedded into the geometrical space of the considering network. The arbitrary coordinate $(p,q)$ assigns the centre position, having the concentration $c(p,q,t)$. The signs, positive and negative point out the relative possibility of the attachment and the leaving of the individuals, respectively. The net effect of a random drift on average concentration of an individual type is a simple sum of the two resultant flux terms of each of the directions where the direction of arrows indicates the directions of the impact of the random drift. This is the schematic presentation of the physical interpretation of the term, $\overline{\nabla}^{2}c = (\frac{\partial^2}{\partial p^2}-\frac{\partial^2}{\partial q^2})c$. The randomness of the term  intuitively understands through the sign convention. The probability to occur each sign at a particular point may assume to be equal to $0.5$ and the factor $0.5$ is part of the $\sigma_{*}$.}
  \label{fig:1}
\end{center}  
\end{figure}

The outcome of the evolution is the result of the combined effects of the two completely independent events: one is the birth-death event, and the other is the random drift event \cite{broom2013game}. In the underlying mechanism of birth-death, the expectation value of transition individuals over each of the population hubs has been calculated with respect of an evolution probability, where the product of the segregation probability and the reproduction-demise probability determines the probability mass function, i.e. the evolution probability. And, finally the expectation of all calculated expectation values of transition individuals over all possible hub sizes measures the evolution rate change  per unit time span due to the birth-death event. The other independent event, the random drift, completely is captured within the probability space, $\Omega$, that being defined by two parameter variables: $p$ is the probability of attachment of a new individual to the network and $q$ is the probability of leaving of an individual from the network (Fig.~\ref{fig:1}). The fractional concentration variables, $x$ and $y$ corresponding to cooperators and defectors respectively, represent the explicit form of the evolutionary dynamics as:
\begin{flalign*}
\frac{\partial{x}}{\partial t}&=a_{3}(\frac{x}{x+y})^3+a_{2}(\frac{x}{x+y})^2+a_{1}(\frac{x}{x+y})+ \sigma_{x}\overline{\nabla}^{2}x, \nonumber \\[\parskip]
\frac{\partial{y}}{\partial t}&=b_{3}(\frac{y}{x+y})^3+b_{2}(\frac{y}{x+y})^2+b_{1}(\frac{y}{x+y})+\sigma_{y} \overline{\nabla}^{2}y, 
\end{flalign*}
\noindent where $a_{*}$s and $b_{*}$s are the mathematical functions of fitness, graph degree, and $\sigma_{*}$s are the coefficients of the random drift. The term concentration means the spatial density here.

\sloppy
The algorithm version of the same dynamical system is examined  not only for optimising our pursuance but also for determining the initial direction of evolutionary dynamics because the direction determination in the randomness pattern by utilising the Monte Carlo simulation procedure \cite{hindersin2019computation} is essential for the analytical version. The components of the Moran process determine the reproduction-demise probabilities; and this determination sets the birth-death rules in the algorithm version. The birth takes place  if
\[\texttt{random()}<\frac{\texttt{fitness of an individual}}{\texttt{maximum fitness of each individual}},\]
where the death of an individual occurs if 
\[\texttt{random()}<\frac{\texttt{number of same type neighbour individuals}}{\texttt{total number of neighbour individuals}}.\] 
\noindent The independent rules are valid until the presence of at least one same type individual of the focal individual. The standard library function, $\texttt{random()}$, gives rise to a random number at each of the iterative steps. Instead of the segregation probability -- defined by the binomial distribution configuration -- the algorithm method uses the manual approach to assess the segregation tendency. Both the probabilities act at the individual level, not at the gene level. In the other event, the random drift rule of an attachment individual set to: $\texttt{random()<p$\cdot\sigma_{*}$}$, while the random drift rule of  a leaving individual set to: $\texttt{random()<q$\cdot\sigma_{*}$}$, where determination of the preferential selection type individual for the leaving or the attaching is needed, which is described by the majority selection rule.

\subsection{The analytical version of the evolutionary dynamics}
\sloppy
The overall system is captured on the belief of the evolution of individual character  not depending on the specific architecture of the network/spatial structure; instead, it is dependent on the average number of neighbours of the focal individual. Each environment region comprises cooperators and defectors, having the normal carrying capacity $\mathcal{N}$. Now, we take $P(i,j,t)$ is the probability that there are $i$ cooperators and $j$ defectors at the time $t$, and denote the increasing and decreasing transition rates of the population from one state to its neighbouring state as $T^{+}_{*}$ and $T^{-}_{*}$ respectively, with appropriate subscripts -- the explanation is that in per unit time, through the rate $T^{+}_{i-1}$  the state $i-1$ switches to the next neighbouring state $i$ while through the rate $T^{-}_{i}$  the state $i$ switches to the next neighbouring state $i-1$. In the time duration $\Delta t$, replacing the simple rate laws by probability laws we have
\begin{flalign}
P(i,j,t+\Delta t)-P(i,j,t)&=T^{+}_{i-1}P(i-1,j,t)+T^{+}_{j-1}P(i,j-1,t) +T^{-}_{i+1}P(i+1,j,t) \nonumber  \\ 
             &  +T^{-}_{j+1}P(i,j+1,t)-(T^{+}_{i}+T^{+}_{j}+T^{-}_{i}+T^{-}_{j})P(i,j,t). 
\end{flalign}
\noindent The system dynamics is generally described by the above master equation \cite{gardiner1985handbook}. In the present frontier, we shall show that based on this probability law, we shall be able to derive the two separate time dependent loci of evolutionary dynamics, one for cooperator group and the other for defector group following the special mathematical tractability, so that the system dynamics can be portrayed explicitly.

\sloppy
Keeping in mind the driving fact of random drift like diffusion \cite{crow1970introduction} directed towards the lower concentration state, in order to convert from the discrete to the continuous state doing slight modification of some appropriate terms in the master equation, the locus of the evolutionary dynamics of cooperation in the presence of the $i$-number of cooperators, on the continuous state, can be extracted as
\begin{flalign}
iP(i,j,t+\Delta t)-iP(i,j,t)&=\{0 \cdot T^{+}_{i-1}P(i-1,j,t)+0 \cdot T^{+}_{j-1} P(i,j-1,t) \nonumber \\[\parskip] 
             &  +iT^{-}_{i+1}P(i+1,j,t)+0 \cdot T^{-}_{j+1}P(i,j+1,t)   \nonumber \\[\parskip]  
             &  -[-iT^{+}_{i}+0 \cdot T^{+}_{j}+2iT^{-}_{i}+0 \cdot T^{-}_{j}]P(i,j,t)\}\Delta t   \nonumber \\[\parskip] 
\implies \sum_{i=0}^{\mathcal{N}}\sum_{j=0}^{\mathcal{N}-i}[iP(i,j,t+\Delta t)-iP(i,j,t)]&=\sum_{i=0}^{\mathcal{N}}\sum_{j=0}^{\mathcal{N}-i}\{iT^{+}_{i}P(i,j,t) 
               -iT^{-}_{i}P(i,j,t) \nonumber \\[\parskip]  
             & +\underbrace{iT^{-}_{i+1}P(i+1,j,t)-iT^{-}_{i}P(i,j,t)}_{\text{random drift effect}}\}\Delta t.   
\end{flalign}
Continuous state takes only one specific state $i$ cooperators at a particular time $t$ regarding a particular hub comprising $i$ cooperators. Due to that reason we eliminate the first term on right hand side by putting zero and other terms related to defector are automatically eliminated. Now the birth-death evolution is defined by the difference between the reproduction term and the demise term, where the reproduction is defined by $-(-iT^{+})$ and after the natural selection it becomes $-(-iT^{+})P(i,j,t)$. The term $-2iT^{-}_{i}$ is the result of two events one for the demise and other for the random drift effect. Clearly, in the random drift effect the transition probability $T^{\pm}_{*}$ depends on the concentrations of the neighbour regions and to set an explicit definition of it would be impossible generally. In this situation, we introduce a new operator, nabla-bar-square, on the probability space which is capable to calculate the random drift effect implicitly. It would be important to mention that in the conversion from the discrete form to the continuous form, the first expression can be interpreted as, a relation resulting in a change in the mathematical expectation value of $i$ cooperators regarding the particular hub -- the concept of population in probability -- while the second expression with the double summations is a relation resulting in a change in the mathematical expectation value of cooperators where the number of cooperators ranging from $0$ to $\mathcal{N}-j$ over the different types of hubs at a particular region. The local hub calculation has been extended to the entire region.  Similarly, using the same master equation the locus of the evolutionary dynamics of defection in the presence of the $j$-number of defectors, on the continuous state, can be extracted as  
\begin{flalign}
jP(i,j,t+\Delta t)-jP(i,j,t)&=\{0 \cdot T^{+}_{i-1}P(i-1,j,t)+ 0 \cdot T^{+}_{j-1} P(i,j-1,t) \nonumber \\[\parskip] 
             &  +0 \cdot T^{-}_{i+1}P(i+1,j,t)+j T^{-}_{j+1}P(i,j+1,t)   \nonumber \\[\parskip]  
             &  -[0 \cdot T^{+}_{i}-jT^{+}_{j}+0 \cdot T^{-}_{i}+2jT^{-}_{j}]P(i,j,t)\}\Delta t   \nonumber \\[\parskip] 
\implies \sum_{j=0}^{\mathcal{N}}\sum_{i=0}^{\mathcal{N}-j}[jP(i,j,t+\Delta t)-jP(i,j,t)]&=\sum_{j=0}^{\mathcal{N}}\sum_{i=0}^{\mathcal{N}-j}\{j T^{+}_{j} P(i,j,t)  
              -jT^{-}_{j}P(i,j,t) \nonumber \\[\parskip]  
& + \underbrace{j T^{-}_{j+1}P(i,j+1,t)-jT^{-}_{j}P(i,j,t)}_{\text{random drift effect}}\}\Delta t.   
\end{flalign}

\noindent Next, we have to calculate each term of each equation with following the specific mathematical tractability in view of the simple natural laws.

\begin{itemize}[leftmargin=*]
\sloppy
\item 
The probability distribution of cooperators and defectors is expanded in a Taylor series at $t$ upto the second order in $\Delta t$ and the fact $x^{\prime}=\sum_{i=0}^{\mathcal{N}}\sum_{j=0}^{\mathcal{N}-i}iP(i,j,t)=\sum_{i=0}^{\mathcal{N}}iP_{i}(i,t)$ yields that
\begin{flalign}
\sum_{i=0}^{\mathcal{N}}\sum_{j=0}^{\mathcal{N}-i}[iP(i,j,t+\Delta t)-iP(i,j,t)]&=\sum_{i=0}^{\mathcal{N}} [\frac{\partial}{\partial t}iP_{i}(i,t)]\Delta t+ \mathcal{O}(\Delta t)^2 \nonumber\\[\parskip]
&=\frac{\partial x^{\prime}}{\partial t}\Delta t+ \mathcal{O}(\Delta t)^2.
\end{flalign}
\noindent Similarly, the probability distribution of cooperators and defectors is expanded in a Taylor series at $t$ upto the second order in $\Delta t$ and the fact $y^{\prime}=\sum_{j=0}^{\mathcal{N}}\sum_{i=0}^{\mathcal{N}-j}jP(i,j,t)=\sum_{j=0}^{\mathcal{N}}jP_{j}(j,t)$ yields that
\begin{flalign}
\sum_{j=0}^{\mathcal{N}}\sum_{i=0}^{\mathcal{N}-j}[jP(i,j,t+\Delta t)-jP(i,j,t)]&=\sum_{j=0}^{\mathcal{N}} [\frac{\partial}{\partial t}jP_{j}(j,t)]\Delta t+ \mathcal{O}(\Delta t)^2 \nonumber\\[\parskip]
&=\frac{\partial y^{\prime}}{\partial t}\Delta t+ \mathcal{O}(\Delta t)^2.
\end{flalign}

\item In this step, to measure the equality, we have to define the segregation probability which is represented by $P$, and the reproduction-demise  probability which is represented by $T$. As according to the order of genetic reproduction actions, the segregation of genes takes place before the natural selection, thus here the segregation probability plays a role of fertility selection which is nothing but the effect of social arrangement of two types of individuals while the natural selection role is played by the reproduction-demise probability. However, in the main master equation the order of the actions of the two probability terms might be different, first we could count the reproduction-demise probability, then the next would be to count the segregation probability, or vice-versa. This is obvious because in the component wise representation of master equations a specific number of individuals is assumed at particular times of $t$ where we confine ourselves in the three consecutive cooperator levels: $(i-1)^{th}$ level, $i^{th}$ level and $(i+1)^{th}$ level as well as in the similar levels for defectors regarding consideration of a hub comprising $i$ cooperators and $j$ defectors. In the present treatise we follow the natural order, first the fertility selection and then after the natural selection at the same social structure, and due to following this order, we are able not only  to carry out the local hub calculation but can also extend the calculation to the whole network structure. One can infer from such pattern of binomial distribution of two different characters of individuals that, the spatial or configurational arrangement or fertility effect is to be perfectly defined by Wright-Fisher transition probability in terms of the expected numbers of cooperators and defectors at a particular time $t$ because the Wright-Fisher transition probability is some extend a more general version of the reproduction rule than the majority selection updating rule, while the reproduction-demise probability is proper fit to the selection component of Moran transition probability. To extend the local hub calculation, we consider the system environment is a connected graph with $\mathcal{N}$ vertices and degree distribution $p(k)$. Therefore, with a note that $z=\sum kp(k)$ and defining the normalised fitness constant $\Gamma_{max}$  in the averagely sense of optimum fitness of an individual, such that $\Gamma_{max}\geq average\{\pi_{CC},\pi_{CD}, \pi_{DC},\pi_{DD}\}$, in the limit of $\Delta t \rightarrow 0 $ we have \let\thefootnote\relax\footnotetext{
Note: If $r.v.= \text{random variable}$, where $r.v.\sim B(k,p_{\scriptscriptstyle{C}})$ and $p_{\scriptscriptstyle{C}}=\frac{x^{\prime}}{x^{\prime}+y^{\prime}}$, $p_{\scriptscriptstyle{D}}=\frac{y^{\prime}}{x^{\prime}+y^{\prime}}$, we have $E(r.v.^{2})=var(r.v.)+E(r.v.)^2=kp_{\scriptscriptstyle{C}}p_{\scriptscriptstyle{D}}+(kp_{\scriptscriptstyle{C}})^2$. Thus, \\
$\sum_{k=1}^{\mathcal{N}-1}p(k)\sum_{i=0}^{k} (\text{expression of}\hspace{1mm} r.v. \sim B(k,p_{\scriptscriptstyle{C}}))\times(\text{other terms}) \\
=\sum_{k=1}^{\mathcal{N}-1}p(k).(kp_{\scriptscriptstyle{C}}p_{\scriptscriptstyle{D}}+(kp_{\scriptscriptstyle{C}})^2)\times(\text{other terms}) \\
=(zp_{\scriptscriptstyle{C}}(1-p_{\scriptscriptstyle{C}})+\sum_{k=1}^{\mathcal{N}-1}k^{2}p(k).p^2_{\scriptscriptstyle{C}})\times(\text{other terms}) \\
=(zp_{\scriptscriptstyle{C}}+(\sigma^2+z^2-z)p^2_{\scriptscriptstyle{C}}).\frac{1}{z\Gamma_{max}}\times(\text{other terms})$}
\begin{flalign}
\lim_{\Delta t \to 0} \frac{1}{\Delta t}\sum_{i=0}^{\mathcal{N}}\sum_{j=0}^{\mathcal{N}-i}[iT^{+}_{i}P(i,j,t)] \Delta t &= \sum_{i=0}^{\mathcal{N}}iT^{+}_{i}P_{i}(i,t)\nonumber \\[\parskip] 
&= \sum_{k=1}^{\mathcal{N}-1} p(k) \sum_{i=0}^{k}iT^{+}_{i}P_{i}(i,t)  \nonumber \\[\parskip]
&= \sum_{k=1}^{\mathcal{N}-1} p(k) \sum_{i=0}^{k}i\frac{k!}{i!(k-i)!} (\frac{x^{\prime}}{x^{\prime}+y^{\prime}})^i (\frac{y^{\prime}}{x^{\prime}+y^{\prime}})^{(k-i)}\nonumber  \nonumber \\[\parskip]
& \times \frac{i}{z \Gamma_{max}}(\frac{x^{\prime}}{x^{\prime}+y^{\prime}}\pi_{CC}+\frac{y^{\prime}}{x^{\prime}+y^{\prime}}\pi_{CD}) \nonumber \\[\parskip] 
&= \{\frac{1}{\Gamma_{max}}(\frac{x^{\prime}}{x^{\prime}+y^{\prime}})+(\frac{\sigma^2+z^2-z}{z \Gamma_{max}})(\frac{x^{\prime}}{x^{\prime}+y^{\prime}})^2\} \nonumber \\[\parskip] 
& \times [\pi_{CD}+(\pi_{CC}-\pi_{CD})(\frac{x^{\prime}}{x^{\prime}+y^{\prime}})], 
\end{flalign}
\noindent $\sigma^{2}$ being the variation of degree distribution. Clearly, the rate $T_{*}^{\pm}$ does not belong to one-step transition probability of Markov chain however, the outcome at any stage depends on the outcome of the previous stage, and the part of the birth-death mechanism is derived by the following formulation: $(\text{binomial coefficient})(\frac{x^{\prime}}{x^{\prime}+y^{\prime}})^{i}. (\frac{y^{\prime}}{x^{\prime}+y^{\prime}})^{N-i}$, where both fractional variables count the effect of vacant places on the  average concentration mechanism. We also incorporate the term probability of degree distribution $ p(k)$ in the calculation but it has no direct effect on the concentration mechanism. It is involved in the calculation as an average degree distribution $z$. Here the term $ p(k)$ supports our claim: $ \mathcal{N}=i+j+\text{vacant places}$, as the act of taking $j=\mathcal{N}-i$ has been rectified by the probability term $ p(k)$ in the calculation steps. $ p(k)$ can take a constant value for a constant value of $k$ however, this consideration is irrelevant to a physical point of view because we already consider vacant vertices here. It means that in this context a random regular graph and a heterogeneous complex network play the same role on the evolutionary dynamics. However, without non-constant $ p(k)$ we can not introduce the term $z$ as well as we can not incorporate the concept of vacant vertices in this analytical method. Also, the local hub calculation can not be extended to the entire region without $p(k)$. It is easily understandable that all perspectives are strongly related to each other. The mechanism of the non-constant value of $p(k)$ and the zero value of $\sigma^{2}$ assures the existence of vacant places in the network topology those places would make the population change. Similarly, in the limit of $\Delta t \rightarrow 0 $, we have 
\begin{flalign}
\lim_{\Delta t \to 0} \frac{1}{\Delta t}\sum_{j=0}^{\mathcal{N}}\sum_{i=0}^{\mathcal{N}-j}[jT^{+}_{j}P(i,j,t)] \Delta t &= \sum_{j=0}^{\mathcal{N}}jT^{+}_{j}P_{j}(j,t)\nonumber \\[\parskip] 
&= \{\frac{1}{\Gamma_{max}}(\frac{y^{\prime}}{x^{\prime}+y^{\prime}})+(\frac{\sigma^2+z^2-z}{z\Gamma_{max}})(\frac{y^{\prime}}{x^{\prime}+y^{\prime}})^2\} \nonumber \\[\parskip] 
& \times [\pi_{DC}+(\pi_{DD}-\pi_{DC})(\frac{y^{\prime}}{x^{\prime}+y^{\prime}})]. 
\end{flalign}
\item In this mathematical part, the reproduction-demise probability is proper fit by the demise component of Moran transition probability where the local hub calculation is extended by the following manner:
\begin{flalign}
\lim_{\Delta t \to 0} \frac{1}{\Delta t}\sum_{i=0}^{\mathcal{N}}\sum_{j=0}^{\mathcal{N}-i}[iT^{-}_{i}P(i,j,t)] \Delta t &= \sum_{i=0}^{\mathcal{N}}iT^{-}_{i}P_{i}(i,t)\nonumber \\[\parskip] 
&= \sum_{k=1}^{\mathcal{N}-1} p(k) \sum_{i=0}^{k}iT^{-}_{i}P_{i}(i,t)  \nonumber \\[\parskip]
&= \sum_{k=1}^{\mathcal{N}-1} p(k) \sum_{i=0}^{k}i\frac{k!}{i!(k-i)!} (\frac{x^{\prime}}{x^{\prime}+y^{\prime}})^i \nonumber  \nonumber \\[\parskip]
& \times (\frac{y^{\prime}}{x^{\prime}+y^{\prime}})^{k-i} \frac{i}{z} \nonumber \\[\parskip] 
&=(\frac{x^{\prime}}{x^{\prime}+y^{\prime}})+(\frac{\sigma^2+z^2-z}{z})(\frac{x^{\prime}}{x^{\prime}+y^{\prime}})^2.    
\end{flalign}
\noindent Similarly, regarding the defector component the evaluated expression is given by  
\begin{flalign}
\lim_{\Delta t \to 0} \frac{1}{\Delta t}\sum_{j=0}^{\mathcal{N}}\sum_{i=0}^{\mathcal{N}-j}[jT^{-}_{j}P(i,j,t)] \Delta t &= \sum_{j=0}^{\mathcal{N}}jT^{-}_{j}P_{j}(j,t)\nonumber \\[\parskip] 
&=(\frac{y^{\prime}}{x^{\prime}+y^{\prime}})+(\frac{\sigma^2+z^2-z}{z})(\frac{y^{\prime}}{x^{\prime}+y^{\prime}})^2.    
\end{flalign}
\item Next, we are interested to measure the possibility  whether the particular population of a focal region could be influenced by out side individuals or not, where  this mathematical part deals with the analytical deriving procedure to capture the effect of the random drift phenomenon to the evolutionary dynamics. Here, we measure the random drift effect on the probability space $\Omega$ which is defined as $\Omega=\{(p=zp^{\prime},q): 0\leq p \leq 1, 0\leq q \leq 1\}$ where $p^\prime$ is the probability of attachment of a new individual at the per degree of a vertex and $q$ is the probability of leaving of an individual at the maximum degree of a vertex; clearly in the sense of probability in the one dimensional space we get: $ q \le \neg q \equiv p$. However, one dimension is not our concern here. Now, introducing a dummy small space length $2\Delta h$ and the fractional concentration variable $x= (\frac{1}{\mathcal{N}})\lim _{\Delta h \to 0}\frac{{x}^\prime}{(2\Delta h)^2}$, examining the essential features of partial conservation law and applying the concept of continuous probability law, intuitively it can be written as   
\begin{flalign}
\lim_{\Delta t \to 0} \frac{1}{\Delta t}\sum_{i=0}^{\mathcal{N}}\sum_{j=0}^{\mathcal{N}-i}[iT^{-}_{i+1}P(i+1,j,t)-iT^{-}_{i}P(i,j,t)] \Delta t & \propto 2\Delta h \cdot \frac{\partial}{\partial p}x(p+\Delta h, q,t)\nonumber \\[\parskip]  
-2\Delta h \cdot \frac{\partial}{\partial p}x(p-\Delta h, q,t)& -2\Delta h \cdot \frac{\partial}{\partial q}x(p, q+\Delta h,t) \nonumber  \\[\parskip] +2\Delta h \cdot \frac{\partial}{\partial q}x(p, q-\Delta h,t) & = (2\Delta h)^2 (\frac{\partial^{2}x}{\partial p^2}-\frac{\partial^{2}x}{\partial q^2}) \nonumber \\[\parskip]
&=(2\Delta h)^2 \overline{\nabla}^{2}x.   
\end{flalign}
\noindent The sign convention has been used on the perception of higher and lower relative possibilities. Similarly, with the fractional concentration variable $y=(\frac{1}{\mathcal{N}})\lim _{\Delta h \to 0 }\frac{{y}^\prime}{(2\Delta h)^2}$, regarding the defector component the evaluated expression is given by 
\begin{flalign}
\lim_{\Delta t \to 0} \frac{1}{\Delta t}\sum_{i=0}^{\mathcal{N}}\sum_{j=0}^{\mathcal{N}-i}[jT^{-}_{j+1}P(i,j+1,t)-jT^{-}_{i}P(i,j,t)] \Delta t & \propto 2\Delta h \cdot \frac{\partial}{\partial p}y(p+\Delta h, q,t) \nonumber \\[\parskip]  
-2\Delta h \cdot \frac{\partial}{\partial p}y(p-\Delta h, q,t) & -2\Delta h \cdot \frac{\partial}{\partial q}y(p, q+\Delta h,t) \nonumber \\[\parskip] +2\Delta h \cdot \frac{\partial}{\partial q}y(p, q-\Delta h,t) & = (2\Delta h)^2 (\frac{\partial^{2}y}{\partial p^2}-\frac{\partial^{2}y}{\partial q^2}) \nonumber \\[\parskip]
&=(2\Delta h)^2 \overline{\nabla}^{2}y.   
\end{flalign}
\end{itemize}
\noindent Therefore, introducing the proportional constants $\sigma_x$, $\sigma_y$ and scale factor $\alpha=\frac{1}{\mathcal{N}}$, the dynamical discrete equations turn out to be the continuous equations having the following forms:
\begin{flalign}
\frac{\partial{x}}{\partial t}&=a_{3}(\frac{x}{x+y})^3+a_{2}(\frac{x}{x+y})^2+a_{1}(\frac{x}{x+y})+ \sigma_x\overline{\nabla}^{2}x ,  \\[\parskip]
\frac{\partial{y}}{\partial t}&=b_{3}(\frac{y}{x+y})^3+b_{2}(\frac{y}{x+y})^2+b_{1}(\frac{y}{x+y})+\sigma_y \overline{\nabla}^{2}y ,
\end{flalign}
\noindent where {\small{$a_{3}=(\frac{\pi_{CC}-\pi_{CD}}{\Gamma_{max}})(\frac{\sigma^2+z^2-z}{z})\alpha$, $a_{2}=(\frac{\sigma^2+z^2-z}{z})(\frac{\pi_{CD}}{\Gamma_{max}}-1)\alpha+(\frac{\pi_{CC}-\pi_{CD}}{\Gamma_{max}})\alpha$, $a_{1}= (\frac{\pi_{CD}}{\Gamma_{max}}-1)\alpha$, $b_{3}=(\frac{\pi_{DD}-\pi_{DC}}{\Gamma_{max}})(\frac{\sigma^2+z^2-z}{z})\alpha$, $b_{2}=(\frac{\sigma^2+z^2-z}{z})(\frac{\pi_{DC}}{\Gamma_{max}}-1)\alpha+(\frac{\pi_{DD}-\pi_{DC}}{\Gamma_{max}})\alpha$, $b_{1}=(\frac{\pi_{DC}}{\Gamma_{max}}-1)\alpha$, $\alpha t= t^{\prime}$}}; and nabla-bar-square measures the net effect of random drift in respect of  the probability space, $\Omega$. As we consider the evolutionary dynamics on network in a finite population, the upper limit of concentration is to be confined by the explicit involvement  of total allocable space $\mathcal{N}$ in both of the above equations; in the change of time scale this involvement totally depends on the specific values of other equation parameters and the trick enables us to measure the whole scenario of evolution in the same scale-up calibration under the constraints of $x\geq0$, $y\geq0$ and $x+y\leq1$, and in the consideration of a periodic boundary condition in $(p,q)$-space. Fig.~\ref{fig:2} shows the  three panel analytical outcomes.

\begin{figure}[t!]
\begin{center}
  \includegraphics[width=0.745\linewidth]{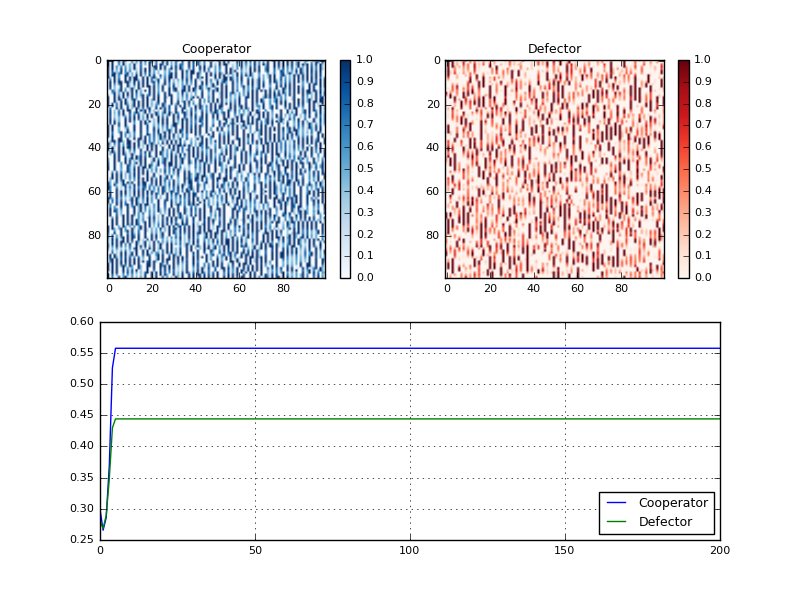}
  \caption{Upper panels: The final concentration distributions of cooperators and defectors over $\Omega$ field of $10,000$ observation points; Lower panel: The analytical version outcomes up to $20,000$ time steps at the point $(p,q)=(0.5,0.4)$. The data corresponding to the base matrix of Fig.~\ref{fig:4} are used.}
  \label{fig:2}
\end{center}  
\end{figure}

\begin{figure}[b!]
\begin{center}
  \includegraphics[width=0.745\linewidth]{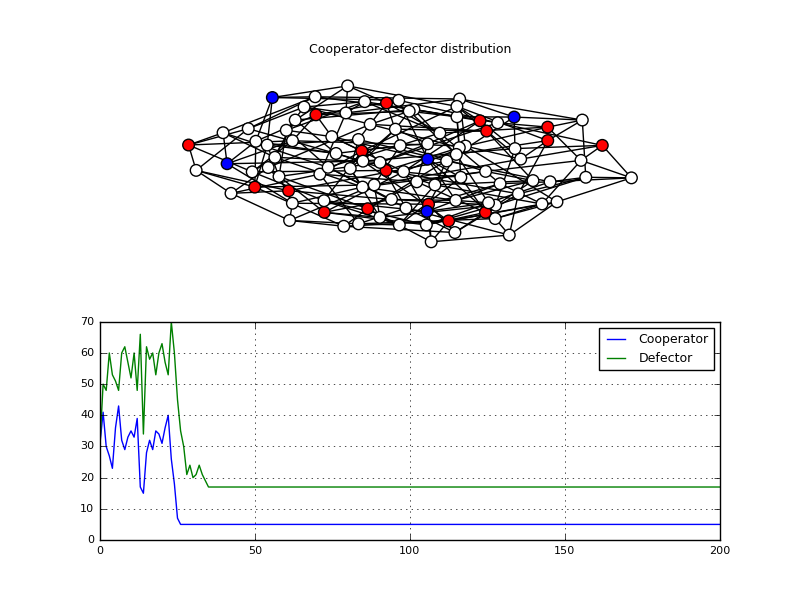}
  \caption{One of the algorithm version outcomes for $\mathcal{N}=100$ up to $20,000$ time steps. Without random drift, the base matrix concerning experimental outcome of the Fig.~\ref{fig:4} is depicted here.}
\label{fig:3}
\end{center}    
\end{figure}

\subsection{The algorithm version of the evolutionary dynamics}
Definitely, the first part of the update process, birth-death, is a slower process than the random drift. Without loss of generality, we assume that its completion time takes $15$ time steps. In the presence of $i$-number of $C$ individuals and $j$-number of $D$ individuals over the neighbourhing vertices, the focal $C$ individual gives birth an offspring by the Moran rule of: $\texttt{random()}<\frac{\texttt{i}\cdot\pi_{CC}+\texttt{j}\cdot\pi_{CD}}{\texttt{z}\cdot\Gamma_{max}}$, where the death of the focal individual occurs by the Moran rule of:  
$\texttt{random()}<\frac{\texttt{i}}{\texttt{z}}$; and similar to a focal $D$ individual. Regarding each of the complement probabilities, an individual will not participate in the evolution process and the system will remain unaltered. The derived algorithm method is a proper discrete version of the continuous evolution version undoubtedly. Under the synchronous consideration, a newborn offspring is placed in a vacant vertex and if it is not possible, then the offspring takes any randomly chosen place in the neighbouring vertices. We obtain a scaling in $\mathcal{O}(z)$. Every individual goes through the evolution process at most once in each generation which has been defined by the $15$ time steps. The time complexity of the whole birth-death computation is of the order $\mathcal{O}(\mathcal{N})$.

\sloppy
On the continuous probability space, in the limit of $\Delta h \rightarrow 0$, the increasing rate of an event of certainty about random drift  with positive sign and the decreasing rate of an event of certainty about random drift with negative sign interpret the randomness at the point $(p,q)$, in the sense that the measurement assumes either positive value, negative value or zero. In the discrete algorithm case, at the point $(p,q)$ mathematically the randomness of net effectiveness of drift on an individual where the possibility of drifted individual type is selected by the majority ratio of the individual type in the neighbourhood area including the focal individual itself is determined separately; an individual attachment rule is of: $\texttt{random()<p$\cdot\sigma_{*}$}$, an individual leaving rule is of: $\texttt{random()<q$\cdot\sigma_{*}$}$, and in the case of complementary probability the rules are unaffected in which the generated random number is outside of the range. The outside influence possibility on the evolution is measured by the random drift along with the local possibility of the birth-death. Both inequalities are considered simultaneously at a particular point where the random drift constant $\sigma_{*}$ controls the speed of the drift. Under the synchronous consideration, a newcomer individual is placed at a vacant vertex and if it is not possible then the newcomer replaces the focal individual. We obtain a scaling in $\mathcal{O}(z)$. Every vertex site goes through the random drift process at most once in each time step. Undoubtedly, here also, both measurement approaches -- continuous and discrete -- are two sides of the same coin. The time complexity of the random drift computation is of the order $\mathcal{O}(\mathcal{N})$. 

We use Python library, $\texttt{NetworkX}$ \cite{hagberg2008exploring}, for studying the algorithmic dynamics on networks \cite{newman2018networks}.  
Fig.~\ref{fig:3} shows the two panel algorithm outcomes.

The main model code is  available in GitHub with identifier \url{https://github.com/bijan0317/cdevolution.git}.

\section{The significance of parameters}
In order to define the term $\frac{0}{0}$ to be equal to $0$ we introduce a bound $K$ such that $0\le x+y < K$ and for such introduction we have $\frac{x}{x+y}=\frac{x}{K}(1+\frac{x+y-K}{K})^{-1}\approx \frac{x}{K^2}(2K-x-y)$, and similarly for $\frac{y}{x+y}$. 

The temporal dynamics possesses  the trivial equilibrium $(0,0)$ and one axial equilibrium at the position $(\tilde{x},0)$ for $a_{1}+a_{2}+a_{3}=0$ when $\Gamma_{max}=b-c$; however, the $(0,\tilde{y})$ is not an equilibrium point since at this position $b_{1}+b_{2}+b_{3}=-(\frac{\sigma^2+z^2-z}{z}+1)$.

In the linearised temporal dynamics the associated Jacobian matrix at the trivial equilibrium is of the form:
\(
   M_{(0,0)}=
  \left( {\begin{array}{cc}
   \frac{2a_{1}}{K} & 0 \\
   0 & \frac{2b_{1}}{K} \\
  \end{array} } \right)
\). If both eigenvalues are negative, the trivial equilibrium is stable.  Introducing, \(
   X=
  \left( {\begin{array}{c}
   x \\
   y \\
  \end{array} } \right) 
\), \(
   \sigma=
  \left( {\begin{array}{c}
   \sigma_{x} \\
   \sigma_{y} \\
  \end{array} } \right) 
\) and considering $\overline{\nabla}^{2}X=-\lambda X $ with $\lambda>0$, at the trivial equilibrium the characteristic matrix of the drift-augmented Jacobian matrix in term of the temporal eigenvalue $\kappa$, is of the form: \(
   \kappa I-M_{(0,0)}+\lambda \sigma =
  \left( {\begin{array}{cc}
   \kappa-\frac{2a_{1}}{K}+\lambda \sigma_{x} & 0 \\
   0 & \kappa-\frac{2b_{1}}{K}+\lambda \sigma_{y} \\
  \end{array} } \right)
\). That is the characteristic equation is given by:  $\kappa^2+d_{1}(\lambda)\kappa+d_{2}(\lambda)=0$, where $d_{1}(\lambda)=(\sigma_{x}+\sigma_{y})\lambda-(\frac{2a_{1}}{K}+\frac{2b_{1}}{K})$ and $d_{2}(\lambda)=\lambda^2-(\frac{2a_{1}}{\sigma_{x}K}+\frac{2b_{1}}{\sigma_{y}K})\lambda+\frac{4a_{1}b_{1}}{\sigma_{x}\sigma_{y}K^2}$. As both $a_{1}= (\frac{\pi_{CD}}{\Gamma_{max}}-1)\alpha$ and $b_{1}=(\frac{\pi_{DC}}{\Gamma_{max}}-1)\alpha$ are negative, the expressions of $d_1 (\lambda)$ and $d_2(\lambda)$ take the positive values. Consequently, each of the temporal eigenvalues has a negative real part and we can say that the occurrence of Turing instability \cite{britton2012essential,nesterenko2017morphogene} is not possible. Also, Hopf bifurcation and zero-eigenvalue bifurcation can not occur here as $d_1 (\lambda)\neq 0$ and $d_2 (\lambda)\neq 0$ in any situation. The steady state is stable to spatially uniform perturbation as $d_1 (0)>0$ and $d_2(0)>0$, i.e. $-(\frac{2a_{1}}{K}+\frac{2b_{1}}{K})>0$ and $\frac{4a_{1}b_{1}}{\sigma_{x}\sigma_{y}K^2}>0$. However, numerical outcomes clearly make aware of us that on the $(p,q)$-space the most of the points exhibit the unstable dynamic character. Therefore, our assumption about the negative value of spatial eigenvalue, $-\lambda$, for each point is not correct -- because we do not know the actual distribution nature of the spatial eigenvalues on the probability space -- where the cooperators will act as activator under the consideration of  $\dot{x}> 0$, i.e. $\frac{2a_{1}}{K}>\lambda \sigma_{x}$ while the defectors will act as inhibitor under the consideration of $\dot{y}<0$, i.e. $\frac{2b_{1}}{K}< \lambda \sigma_{y}$; that means the cooperators stimulate their own production and the production of defectors while the defectors inhibit their own production and the production of cooperators. Here, the positive spatial eigenvalue indicates that the individuals enter into the considering region. 

In the linearised temporal dynamics the associated Jacobian matrix at the axial equilibrium is of the form:
\(
   M_{(\tilde{x},0)}=
  \left( {\begin{array}{cc}
   0 & -\frac{1}{\tilde{x}}(3a_{3}+2a_{2}+a_{1}) \\
   0 & \frac{b_{1}}{\tilde{x}} \\
  \end{array} } \right)
\). As one of the eigenvalues is zero and the other is negative, thus the axial equilibrium is non-hyperbolic where the Hartman-Grobman Theorem is not applicable. The characteristic equation related to the temporal eigenvalue $\kappa$, is of the form: $\kappa^2+d_{1}(\lambda)\kappa+d_{2}(\lambda)=0$, where $d_{1}(\lambda)=(\sigma_{x}+\sigma_{y})\lambda-\frac{b_{1}}{\tilde{x}}$ and $d_{2}(\lambda)=\sigma_{x} \sigma_{y}\lambda^2-(\frac{b_{1}}{\tilde{x}}\sigma_{x} )\lambda $. We see that Hopf, Turing and zero-eigenvalue bifurcations can not occur. For $d_{1}(0)= - \frac{b_{1}}{\tilde{x}}$ and $d_{2}(0)=0$, we get $\kappa=0$ and $\kappa= \frac{b_{1}}{\tilde{x}}$. The canonical form of the temporal system to spatially uniform perturbation can be written as: $\dot{x}= 0$, $\dot{y}= \frac{b_{1}}{\tilde{x}}$. Its solution with $x(0)=x_{0}$, $y(0)=y_{0}$ is given by $x(t)=x_{0}$, $y(t)=y_{0}e^{{\frac{b_{1}}{\tilde{x}}}t}$. As $\frac{b_{1}}{\tilde{x}}<0$, all the points on the $x$-axis are stable equilibrium points. Now if $\Gamma_{max}=\pi_{DC}$, then the solution is given by $x(t)=x_{0}$,
$y(t)=y_{0}$, i.e. the solution does not depend on generation and is depending on initial conditions. Thus, $(\tilde{x},0)$ is a stable equilibrium point.   

It is naturally  expected that before both types of individuals go extinct or in their internal position they can form a non-hyperbolic equilibrium state which defines the  non-trivial equilibrium, but to find out an internal equilibrium in the system is not possible in general by following the formal approach to determine the equilibrium points as the conservation principle is not directly used here; however we can obtain internal equilibrium by imposing  constraints through the stepwise procedure. If $K\approx 0$ and $0\le \frac{x+y-K}{K} < 1$, then we can determine a non-hyperbolic equilibrium point $(\tilde{x}, \tilde{y})$ such that $\tilde{x}+\tilde{y}=2K+error$. The characteristic equation related to the temporal eigenvalue $\kappa$, is of the form: $\kappa^2+d_{1}(\lambda)\kappa+d_{2}(\lambda)=0$, where $d_{1}(\lambda)=(\sigma_{x}+\sigma_{y})\lambda+(\frac{a_{1}\tilde{x}}{K^2}+\frac{b_{1}\tilde{y}}{K^2})$ and $d_{2}(\lambda)=\sigma_{x} \sigma_{y}\lambda^2+(\frac{a_{1}\tilde{x}}{K^2}\sigma_{y}+\frac{b_{1}\tilde{y}}{K^2}\sigma_{x})\lambda $. As $d_{1}(\lambda)>0$, $d_{2}(\lambda)>0$, the system does not exhibit Hopf, Turing and zero-eigenvalue bifurcations. Also, we note that $d_{1}(0)=(\frac{a_{1}\tilde{x}}{K^2}+\frac{b_{1}\tilde{y}}{K^2})<0$ and $d_{2}(0)=0$. Thus, $(\tilde{x}, \tilde{y})$ is an unstable non-hyperbolic equilibrium point. We know that the dynamics around non-hyperbolic points is not robust. Adding non-linearities terms and/or changing parameters may lead to strong unpredictability.

If the dynamical system is written as $\dot{x}= P(x,y)$, $\dot{y}= Q(x,y)$, then it be easily verified that $\frac{\partial}{\partial x}(\psi P)+ \frac{\partial}{\partial y}(\psi Q)\neq 0$ in the simply connected region where $\psi$ being a weighted factor; and according to Bendixson's criteria, there are no limit cycles which are the relations of the typical oscillatory behaviours between the interacting species. Also, it is well known, Hopf bifurcation gives rise to limit cycles in many nonlinear systems but as the system does not exhibit Hopf bifurcation we can rule out the concept of existence of limit cycles in the plane. The centre manifold theorem is a model reduction technique for determining the local asymptotic stability of non-hyperbolic equilibrium. However, here we are unable to define the centre manifold $W^{c}$ because variables $x$ and $y$ are not directly related through a function in the constraint of $|x|<\delta $. As the model based on the random distribution structure, except the stable manifold $W^{s}$, the unstable and centre invariant manifolds $W^{u}$ and $W^{c}$ are not possible to define in a formal way. We need more information for determining the proper nature of subspaces $E^{s}$, $E^{u}$, $E^{c}$, corresponding to the span of the generalised eigenvectors.

\begin{figure}[t!]
\makebox[\textwidth][c]{\includegraphics[width=1.12\linewidth]{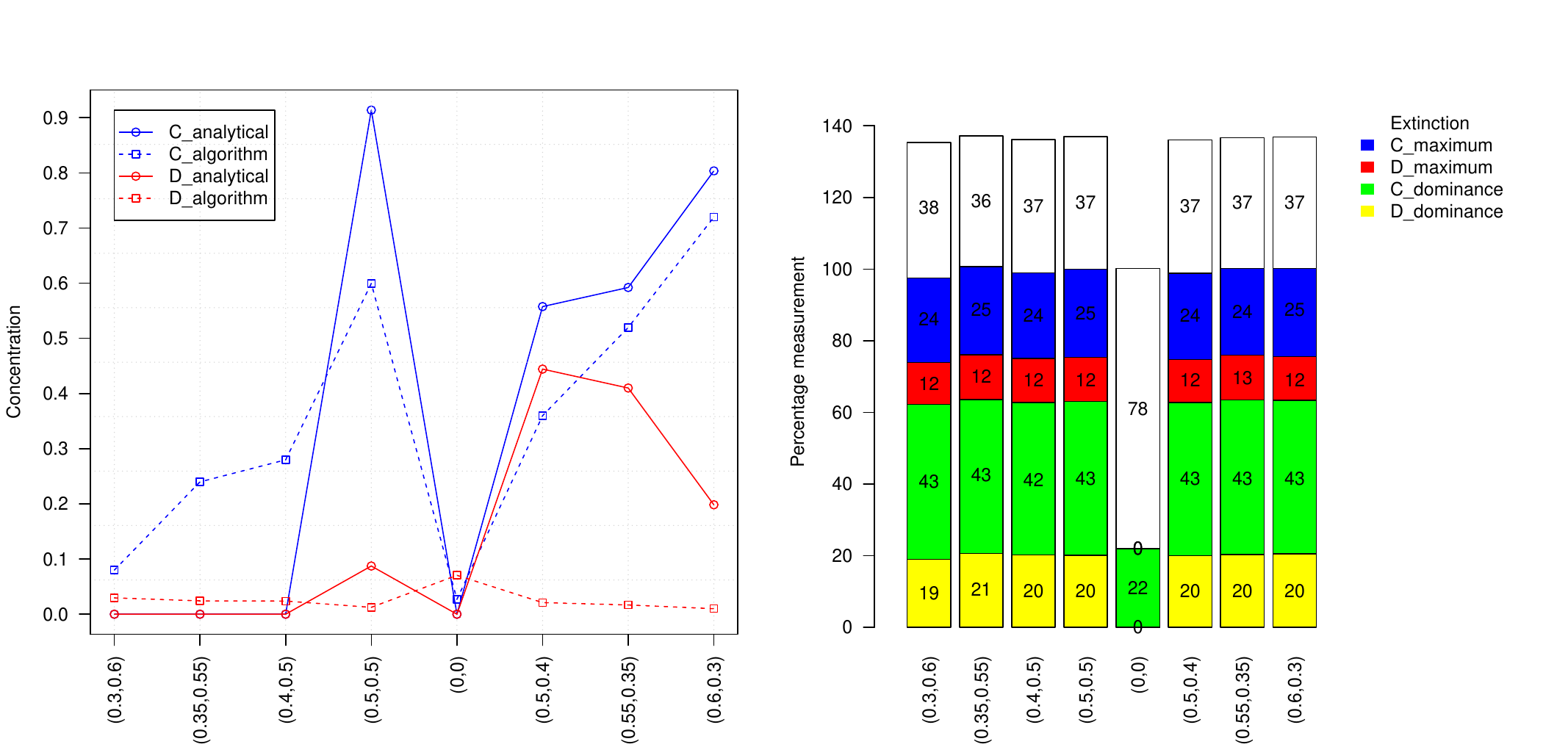}}
  \caption{Left panel: The qualitative comparison between the evolutionary outcomes of the analytical procedure and the algorithm method; Right panel: The percentage measurement based on the $10,000$ observations related to evolutionary outcome at each grid site of the analytical procedure. The outcomes  after $20,000$ time steps with a generation interval of $0.02$ time scale are examined in consideration of  maximum fitness, $0.77$, and of fitness payoffs: $\pi_{CC}=1.5$, $\pi_{CD}=-0.3$, $\pi_{DC}=1.8$, $\pi_{DD}=0$. The number of allocable vertices and the degree of the graph respectively are $1000$ and $50$ where the random drift coefficient in the analytical procedure and the algorithm method respectively is $10^{-5}$ and $0.3\times 10^{-1}$.}
  \label{fig:4}
\end{figure}

\section{Method outline and outcome}
Unlike the previous studies~\cite{lieberman2005evolutionary,zukewich2013consolidating,allen2017evolutionary}, the two distinct evolutionary mechanisms have been described on the random regular graph of degree $z$. In each of the initial stages at the attempt of equal distribution, it is tried to randomly assign the cooperators, defectors and vacant places over the vertices of the graph so that none of the types get an  additional advantage in their evolutionary process. The synchronous model update combines the two steps: one is the birth-death update where the pair do not fully depend on each other, and the second one is the random drift step where the preferential type chooses on the concept: the like-minded individuals would always attract to each other and the act is operated by the majority selection rule. Without loss of generality, we take $15$ time steps as one generation  in which either the birth or the death occurs at most once. However, according to the normal law the birth-death step gets more preference than the random drift evolutionary step. The experimental data of the algorithm method are collected by averaging over the outcome values for running the program up to $20,000$ time steps in the $25$ different graph structure realisations. 

Naturally, we select the same initial averaging concentration value in the analytical procedure for the determination of evolutionary dynamics at the same observation point of $(p,q)$, in the Fig.~\ref{fig:2} the point is $(0.5,0.4)$, where the initial concentrations of other observation points assign by the uniform distribution law, applied in the range $(0,0.3)$. The analytical procedure acts as a correcter tool and run it several times to correct our algorithmic prediction. Both the periodical condition and the bounded condition: $x+y\leq1$, take into the consideration on the physical space of $100\times 100$ locations; each location as a point linearly maps to $\Omega$ space. In the geometrical sense, we consider $10,000$ connected regions where our interest to analysis the comparative study of the evolutionary dynamics at the focal region comprising the different types of hubs.

The outcomes of both evolutionary mechanisms related to the analytics and the algorithm, ensure the ubiquitous character of cooperation in the presence of the random drift. The $(p,q)$ data-range represents the comparative outcomes (the Left panel of Fig.~\ref{fig:4}). Both independent quantities $p$ and $q$ have been considered as  random variables in respect to time scale, while in another turn where time scale is insignificant those quantities act as  non random parameter variables. For different aspects, an arbitrary variable can have such dual character simultaneously. In the model development, we use this intuitive sense. Specifically, a comparative study has been performed on the eight values of the pair $(p,q)$. The evolutionary dynamics derived by the analytical process and the evolutionary dynamics derived by the algorithm method  are not quantitatively equal, however, they are qualitatively equivalent. Its prime reason is: natural laws completely define the analytical process, whereas the algorithm method is based on few known rules. Also, the analytical process is of an implicit nature, whereas the  algorithm method is of an explicit nature. One of the reasons can be clarified as, the random drift effect in the analytical process depends on the spatial densities of the neighbour regions -- in other words, depending on concentrations of different individual types of the same generation,  while the random drift effect in the algorithm method directly depends on the value of pair $(p,q)$. Due to that we notice a deviation between the results of the two approaches on the values of the pair $(p,q)$. However, these deviations can be minimised manually if we increase the number of analytical observations. We place the two approaches side by side in order to mainly show the dominant enhancement of cooperation over defection to be qualitatively independent of environmental scenario, where we have tried to minimise the deviation between the cooperation values. 

\begin{figure}[t!]
\makebox[\textwidth][c]{\includegraphics[width=1.12\linewidth]{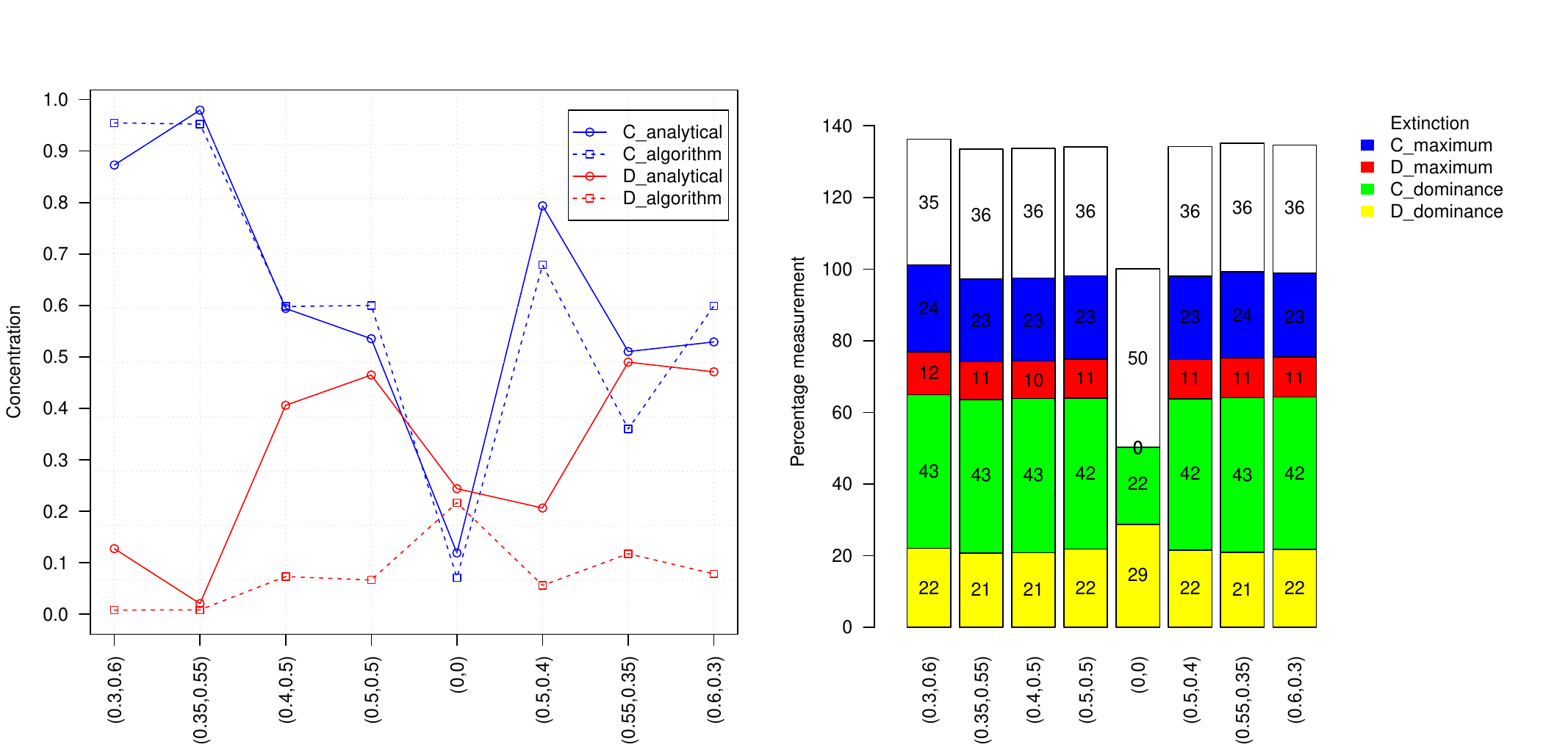}}
  \caption{Base matrix game: The outcomes  after $20,000$ time steps with a generation interval of $0.002$ time scale are examined provided the number of   allocable  vertices and the degree of the graph being $100$ and $5$, respectively.}
  \label{fig:5}
\end{figure}

\begin{figure}[b!]
\makebox[\textwidth][c]{\includegraphics[width=1.12\linewidth]{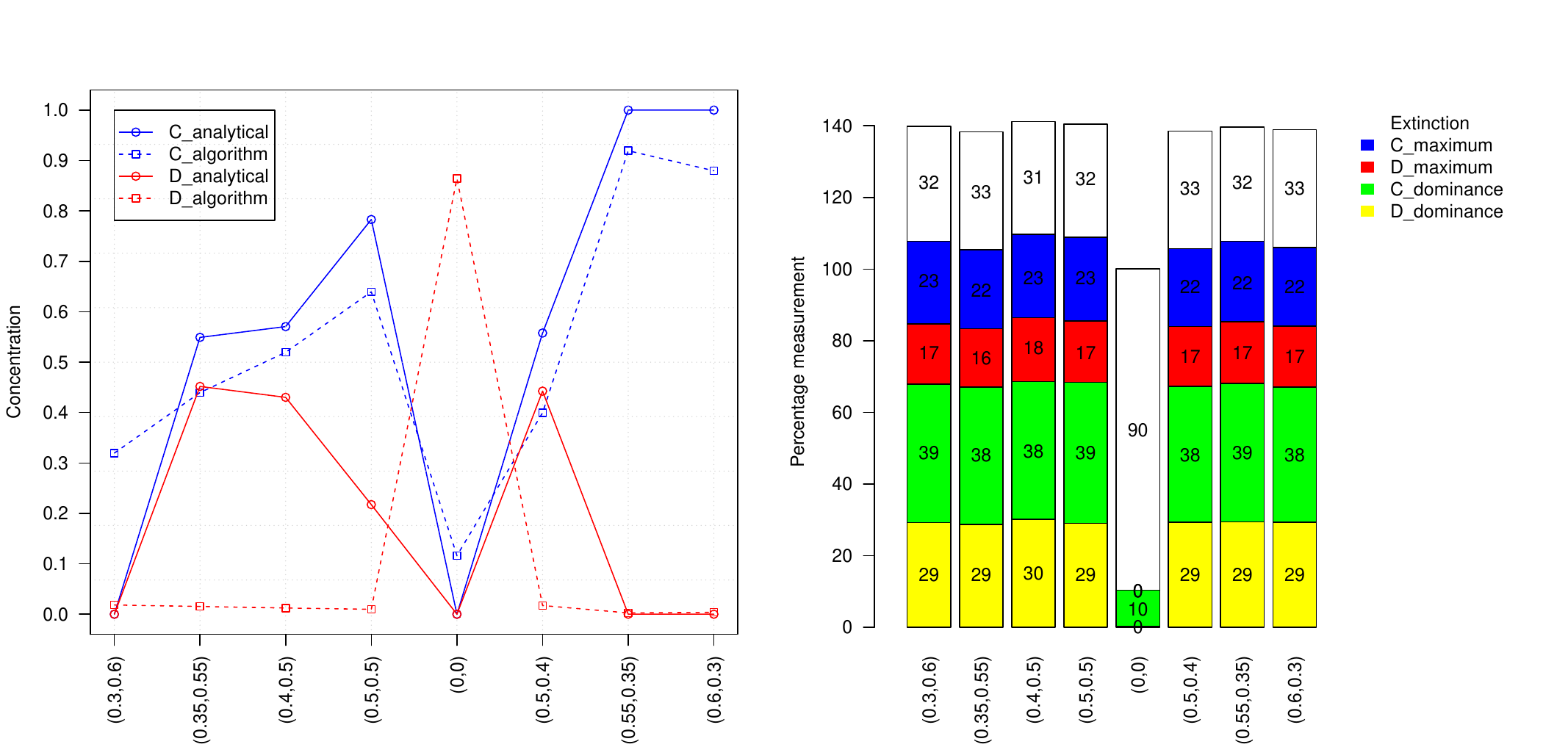}}
  \caption{Prisoner's dilemma game: The outcomes  after $20,000$ time steps with a generation interval of $0.02$ time scale are examined in the consideration of  maximum fitness, $1.145$, and of fitness payoffs: $\pi_{CC}=1.5$, $\pi_{CD}=0.3$, $\pi_{DC}=2.2$, $\pi_{DD}=0.5$. The number of allocable vertices and the degree of the graph respectively are $1000$ and $50$ where the random drift coefficient in the analytical procedure and the algorithm method respectively is $10^{-5}$ and $0.5\times 10^{-1}$.}
  \label{fig:6}
\end{figure}  

In the Right panel of Fig.~\ref{fig:4}, the five dependent observables in the scale of $0$ (or extinction) to $1$ (or maximum -- the state of all $C$ or all $D$) are introduced. The observables are self-explanatory and have been counted on $10,000$ observation points. Clearly, the Right panel of a graphical representation of the experimental data of the analytical procedure represents the percentage measurement of the observation points of the each of the observables; it means that how many times we observe a specific observable having a specific character in the $10,000$ observation points. For example, if we ignore the random drift effect, i.e. at the point $(p,q)=(0,0)$ we see that $22\%$ locations exhibit the character of $\mbox{C\_dominance}$, $29\%$ locations exhibit the character of $\mbox{D\_dominance}$, in the $50\%$ locations population be totally extinct, and no locations do exhibit either the character having all $C$ or the character having all $D$, i.e. both values of $\mbox{C\_maximum}$ and $\mbox{D\_maximum}$ are zero. To get a clarification we also mention the reason that as all observables are not independent to each other, along the ordinate the percentage measure scale is calibrated from $0$ to $140$. The experiment is carried out over three game classes, namely, Dominance game, Coexistence game and Coordination game with either $\mathcal{N}=100$ or $\mathcal{N}=1000$, where we uniformly set initial conditions so that either of types do not get an initial advantage in every possible position, but the uniform distribution of the random initial conditions has its own limitation \cite{perc2017statistical}. According to expectation in the mixed population where defectors have higher fitness than cooperators natural selection favours defectors. However, in the absence of random drift we even observe the phenomenon of the defector extinction. This observation has a simple interpretation. In the intratype competition, the defector fitness payoff is either zero or very minimum, and as a consequence, in the presence of a low number of cooperators the survival capacity of defectors markedly reduces and ultimately they become extinct in the most of the possible situations. Evidently, because of the implicit nature, except for the value $(p,q)=(0, 0)$, in the graphical presentation of the experimental data of the analytical process it seems that the amount of cooperators and defectors is almost the same for different values of $(p,q)$ in a specific game class. Based on data analysis at the value of $(p,q)=(0,0)$ regarding the three game classes, we can reasonably infer that the intratype fitness of individual species -- $\pi_{CC}$ and $\pi_{DD}$ --  is the key factor for not only surviving, but thriving. 

Next, the brief experiment records of the game specific outcomes have been reported in the following subsections. We include the implementation approach of model analysis to the extended cooperation-defection scenarios defined by the four different payoff matrices corresponding to the three game classes. As there are vacant places on the network, the model belongs to the class of dynamical network models in which the network topology is not fixed throughout time.

\subsection{Base matrix game}
In what extent the evolutionary process having significantly been influenced by total number of individuals is prominently notice in Fig.~\ref{fig:5} at the absence of random drift.
One can  conclude here that the total number of individuals controls the rate of evolution. Another base matrix game is examined in the consideration of fitness payoffs: $\pi_{CC}=1.5$, $\pi_{CD}=-0.7$, $\pi_{DC}=2.2$, $\pi_{DD}=0.5$, where the random drift coefficient set to $0.5\times 10^{-1}$ in the  algorithm method. For $\mathcal{N}=100$, observable values with and without random drift respectively are: $38\le \mbox{C\_dominance} \le 40$, $23\le \mbox{D\_dominance} \le 25$ and  $\mbox{C\_dominance}=12$, $\mbox{D\_dominance}=17$ while for $\mathcal{N}=1000$ : $39\le \mbox{C\_dominance} \le 40$, $23\le \mbox{D\_dominance} \le 24$ and $\mbox{C\_dominance}=13$, $\mbox{D\_dominance}=0$.

\subsection{Prisoner's dilemma game}

Prisoner's dilemma game is an example of D-dominance game class of very minimum defector fitness payoff in intratype competition, see Fig.~\ref{fig:6}. For $\mathcal{N}=100$, observable values with and without random drift respectively are: $38\le \mbox{C\_dominance} \le 39$, $29\le \mbox{D\_dominance} \le 31$ and  $\mbox{C\_dominance}=14$, $\mbox{D\_dominance}=54$.

\subsection{Snowdrift game}
\begin{figure}[t!]
\makebox[\textwidth][c]{\includegraphics[width=1.12\linewidth]{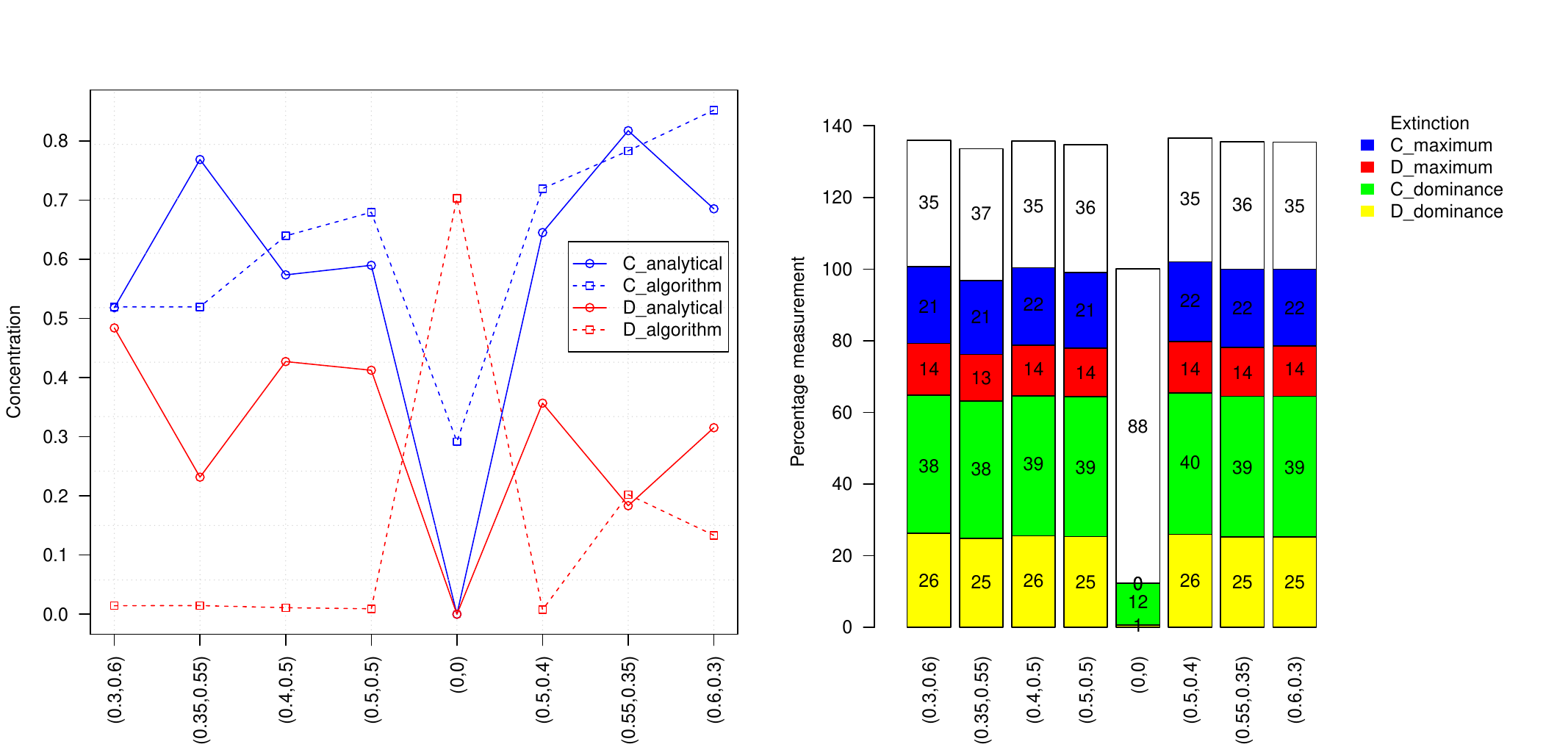}}
  \caption{Snowdrift game: The outcomes  after $20,000$ time steps with a generation interval of $0.02$ time scale are examined in the consideration of  maximum fitness, $1.145$, and of fitness payoffs: $\pi_{CC}=1.5$, $\pi_{CD}=0.5$, $\pi_{DC}=2.2$, $\pi_{DD}=0.3$. The number of allocable vertices and the degree of the graph respectively are $1000$ and $50$ where the random drift coefficient in the analytical procedure and the algorithm method respectively is $10^{-5}$ and $0.5\times 10^{-1}$.}
  \label{fig:7}
\end{figure}

Snowdrift game is an example of coexistance game class of relatively high defector fitness payoff in intertype competition, see Fig.~\ref{fig:7}. For $\mathcal{N}=100$, observable values with and without random drift respectively are: $39\le \mbox{C\_dominance} \le 40$, $25\le \mbox{D\_dominance} \le 26$ and  $\mbox{C\_dominance}=20$, $\mbox{D\_dominance}=54$.

\subsection{Stag-hunt game}

\begin{figure}[b!]
\makebox[\textwidth][c]{\includegraphics[width=1.12\linewidth]{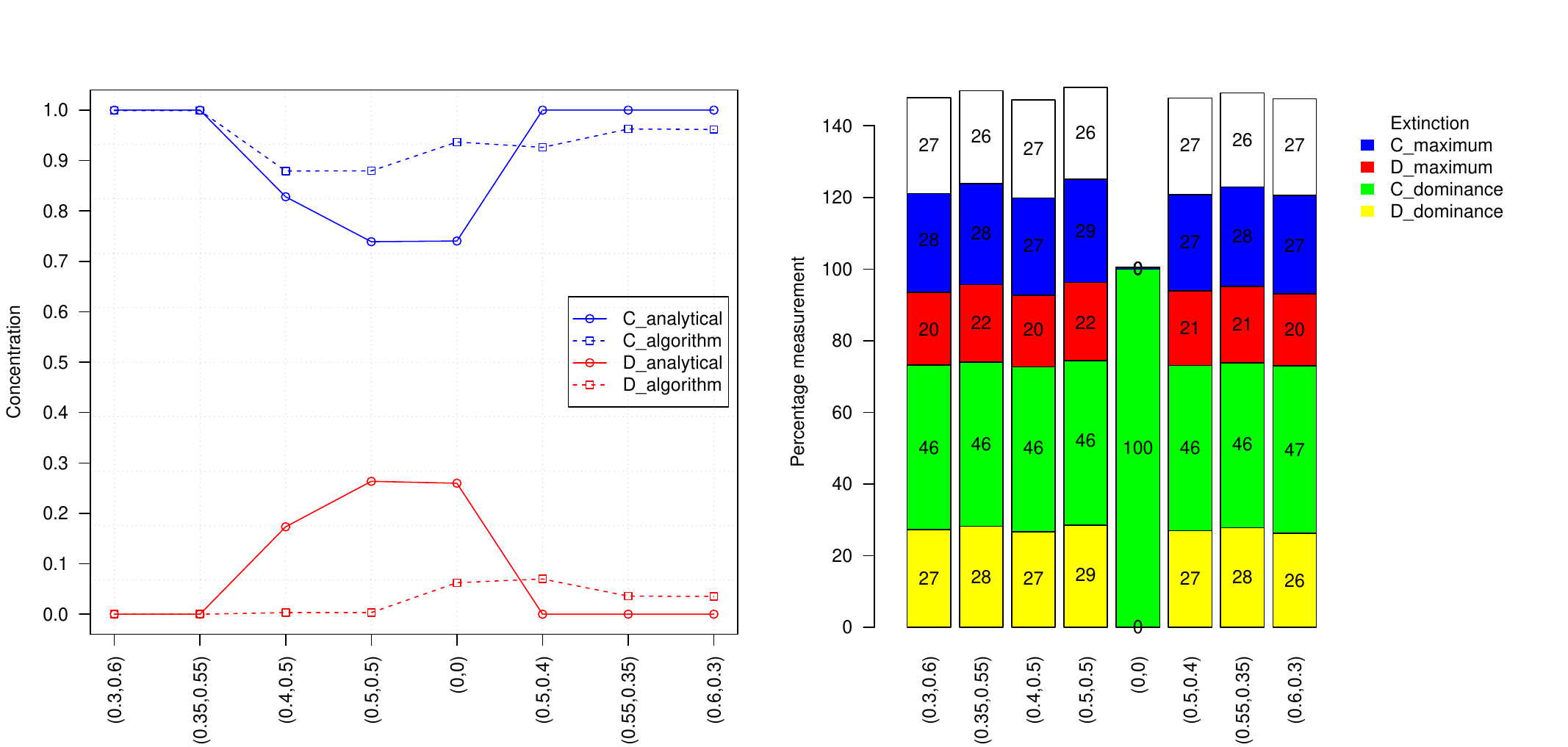}}
  \caption{Stag-hunt game: The outcomes  after $20,000$ time steps with a generation interval of $0.02$ time scale are examined in the consideration of  maximum fitness, $1.145$, and of fitness payoffs: $\pi_{CC}=2.2$, $\pi_{CD}=0.3$, $\pi_{DC}=1.5$, $\pi_{DD}=0.5$. The number of allocable vertices and the degree of the graph respectively are $1000$ and $50$ where the random drift coefficient in the analytical procedure and the algorithm method respectively is $10^{-5}$ and $0.3\times 10^{-1}$.}
  \label{fig:8}
\end{figure}

Stag-hunt game is an example of coordination game class of relatively high cooperator fitness payoff in intratype competition, see Fig.~\ref{fig:8}. For $\mathcal{N}=100$, observable values with and without random drift respectively are: $46\le \mbox{C\_dominance} \le 47$, $27\le \mbox{D\_dominance} \le 29$ and  $\mbox{C\_dominance}=100$, $\mbox{D\_dominance}=0$.

\section{Conclusion}

The evolution is the consequence of gene variation. One of the factors for the gene variation is the fitness of the individual which is the combination of the fertility parameter and the viability parameter. The gene variation also occurs due to the random drift effect. The random drift is the result of the random change in physiological pattern, social behaviour, ecology, etc. Thus, determination of the specific reason for the random drift can be difficult. In the present research paper, the mechanism of cooperation evolution and the mechanism of defection evolution about genetic variation are defined by two independent events: the birth-death event and the random drift event, where in the analytical procedure the nabla-bar-square operator  measures  the drift effect. The effect of the operator is determined on a probability space. Natural laws completely define the analytical process of the derivation of the evolutionary dynamics, whereas the algorithm method of the derivation of the evolutionary dynamics is based on few known rules, and over the initial random distribution of cooperators and defectors, both approaches exhibit that the dominant character of cooperation over defection remains unaffected on any environmental circumstances. Thus, according to the applying method and the applying process to the collecting information, in our experiment our result  firmly demonstrates that natural cooperation is a principle of evolution \cite{nowak2006five}, as principles are ideas based on scientific rules and natural laws.  

The visualised experimental data along with numerical data in the additive record, measure the efficiency of the random drift at sustaining cooperation. In fact, the unbiased attachment rule lays down the equally probable situation for attachment of two types. However, attachment of relatively more defectors than the cooperators leads to extinction, while the attachment of cooperators increases the survival capacity of individuals, irrespective of their behaviour. As extinction environment does not have any effect on the plausible survey result of individual promoting, the evolution track always revealed the fact of cooperation emerging and enhance in front of our eyes. This is why the cooperative behaviour is very natural to observe in various forms of our life system. The explanation does not violate our belief that under the neutral role play, natural environment always tries to retain the environment of coexistence traits \cite{reichenbach2007mobility,maciejewski2014evolutionary,sarkar2018moran} -- the state of biodiversity.  
  
Of particular note is on the ranges of the model parameters the value of $b_{1}+b_{2}+b_{3}$ which is never equal to zero; consequently, we can not consider the point $(0,\tilde{y})$ as an equilibrium point, and it provides us the reason for the arising question about the chance of survival of defectors without cooperator while $a_{1}+a_{2}+a_{3}$ equal to zero can be evaluated, means that $(\tilde{x},0)$ can be an equilibrium point in the evolutionary process. Furthermore, one of the explanations of pattern formation through Turing instability \cite{nakao2010turing} is completely ruled out in this mathematical formalism since the priori assumption $\overline{\nabla}^{2}\bigl( \begin{smallmatrix}x\\y\end{smallmatrix}\bigr) =-\lambda \bigl( \begin{smallmatrix}x\\y\end{smallmatrix}\bigr) $ with spatial eigenvalue, $\lambda$, greater than zero is not acceptable for every point on  $\Omega$ field. Even though if we accept this in some compromise sense probability to occur pattern formation will be almost zero.

Here, in the many circumstances, as fitnesses of individuals depend on the frequencies of cooperators and defectors, the Fisher's fundamental theorem of natural selection which states that average fitness is increased or remains the same under constant selection does not hold. Concerning the same topic it is worth mentioning that the Hardy-Weinberg law \cite{ewens2012mathematical} approved  by this mathematical formalism is to be interpreted  by, $\dot{x}=\dot{y}=0$, at the set payoff values equal to a constant value. 

A comparison has been made between the algorithm method and the analytical procedure due to the intention to demonstrate the natural cooperation characterised by a principle rather than a specific rule of an environmental scenario. For conducting the experiments we consider all possible interrelations among fitness values of two different types of individuals where one can realise that algorithm version is good for a better prediction to point out the initial direction and to find out the final outcome of evolutionary fate while the analytical procedure being a good tracer of the intermediate evolutionary path. The evolutionary path which governs by many of the unpredictable characters of nature to be explained by  preassigned strict rules (or randomly ever-changing set of interactors) over a long time is some extend unrealistic while the analytic one entailed by the average sense can produce closed realistic evolutionary results. Because of this intuitive belief, we use the algorithm version as a predictor tracer and the analytical version as a corrector tracer of an evolutionary path. We anticipate our purely mathematical treatment to be a better choice than the agent-based method~\cite{adami2016evolutionary} to simulate the complexity of heterogeneous populations. The simple design structure, in particular the nabla-bar-square will have great potential for examining a topological dynamical system in the higher dimension, and to gain the knowledge we have to walk the extra mile. 

In this article, we show that the evolution about genetic variation, combining the effects of the social evolution of the birth-death on network structure and of the random drift, gives the final verdict on the selection of the individual type. The social evolution, defined also by average birth-death effect in term of the pure replicator equation, depends only on constant fitness values of individuals and reasonably its outcomes are nothing but the normal display of the linear impact of the payoff-to-fitness mapping and consequently it might be failing to attain a generality. The one of the factors considering for cooperation enhancement is the effect of a social network structure. However, it would be expected for unbiased distribution of individuals, both types of individual would get an equal benefit from the social structure effect. Therefore, one simple conclusion is that over a long evolutionary process the feature of the intratype interaction acts the key role in cooperation enhancement and defection hindrance where the driving force of lending a helping hand of themselves provides the instrumental clues to their future evolutionary fate. 

\section{Research data} \label{C}
The main model code is  available in GitHub with identifier \url{https://github.com/bijan0317/cdevolution.git}.

\section{Acknowledgment} \label{D}
I would like to thank all anonymous reviewers for their valuable comments and constructive suggestions on the standard of the presentation and explanation of the manuscript. The constructive suggestions and healthy criticism about the merit of the research paper are always welcome. 


\bibliographystyle{unsrt}


\end{document}